\documentclass[11pt]{article}

\usepackage{latexsym,amssymb}
\usepackage{cite,epsf}

\newcommand{\ket}[1]{|#1\rangle} 
\newcommand{\bra}[1]{\langle #1|}
\newcommand{\C}[0]{{\mathbb{C}}}
\newcommand{\R}[0]{{\mathbb{R}}}
\newcommand{\N}[0]{{\mathbb{N}}}
\newcommand{\Z}[0]{{\mathbb{Z}}}
\newcommand{\onemat}[0]{\mbox{\bf 1}}
\newcommand{\GL}[0]{{\rm GL}}
\newcommand{\nix}[1]{{}}

\newcommand{\DFT}[0]{{\rm DFT}}
\newcommand{\qed}[0]{\hfill $\Box$}

\newtheorem{theorem}{Theorem}
\newtheorem{definition}[theorem]{Definition}

\newtheorem{example}[theorem]{Example}

\newtheorem{algorithm}[theorem]{Algorithm}

\title{\Large \textbf{
Implementation of Group-Covariant POVMs\\ by Orthogonal Measurements}}

\author{Thomas Decker, Dominik Janzing
\\ IAKS Prof. Beth, Arbeitsgruppe Quantum Computing,
Universit{\"a}t Karlsruhe,\\ Am Fasanengarten 5,
D-76\,131 Karlsruhe, Germany\\
\texttt{\{decker,janzing\}@ira.uka.de}\\[2ex]
\and
Martin R{\"o}tteler\\
Institute for Quantum Computing, University of Waterloo\\ 
Waterloo, Ontario, Canada, N2L 3G1\\
\texttt{mroetteler\symbol{64}iqc.ca}
}

\date{July 7, 2004}

\begin{document}

\maketitle

\begin{abstract}

We consider group-covariant positive operator valued measures (POVMs)
on a finite dimensional quantum system. Following Neumark's
theorem a POVM can be implemented by an orthogonal measurement on a
larger system. Accordingly, our goal is to find an implementation of a
given group-covariant POVM by a quantum circuit using its
symmetry. Based on representation theory of
the symmetry group we develop a general approach for the
implementation of 
group-covariant POVMs which consist of rank-one operators. The 
construction relies on a
method to decompose matrices that intertwine two
representations of a finite group. We give several examples for which
the resulting quantum circuits are efficient. In particular, we obtain
efficient quantum circuits for a class of POVMs generated by Weyl-Heisenberg
groups. These circuits allow to implement an approximative
simultaneous measurement of the position and crystal momentum of a
particle moving on a cyclic chain.

\end{abstract}

%
%

\section{Introduction}

General measurements of quantum systems are described by positive
operator-valued measures (POVMs) \cite{Holevo:82,Helstrom:76}. For
several optimality criteria the use of POVMs can be advantageous as
compared to projector valued measurements. This is true, e.\,g., for
the mean square error, the minimum probability of error \cite{Eldar},
and the mutual information \cite{Sasaki}. POVMs are more flexible
than orthogonal von Neumann measurements and can consist of finite as
well as of an infinite number of elements. An example for the latter
is given in \cite{SpinAriano} where a POVM for measuring the spin
direction is proposed. Here we restrict our attention to the finite case
where a POVM is described by a set of positive operators which sum up to
the identity. Such a POVM is called group-covariant if the
set is invariant under the action of a group.
The example of POVMs for the Weyl-Heisenberg groups
as well as the example in \cite{SpinAriano}
show that POVMs are needed to describe phenomenologically
the mesoscopic scale of quantum systems. They allow
{\it approximatively} simultaneous measurements of quantum observables
which are actually incompatible. For instance, 
the classical phase space of a particle 
can be approximatively reproduced by
simultaneous measurements of momentum and position. 
Descriptions of quantum particles which have strong analogy 
to the classical phase space are helpful to understand the relations
between the classical and the quantum world
\cite{GJK+:96}. Also for several other tasks in
quantum information processing
the implementation of POVMs is of interest
\cite{Aravind,cabello,Fuchs}.

Neumark's theorem \cite{Peres:90,Peres} states that in principle every
POVM can be implemented by an orthogonal measurement of the joint
system consisting of the system and an ancilla system. However, the
orthogonal measurement required by this construction may not be a
``natural'' observable of the joint system. One may need an
additional unitary transform to obtain a reduction to a more natural
observable which henceforth will be called the measurement in the
computational basis of the quantum system.

Therefore, the question arises how to actually implement a POVM in terms
of a quantum circuit which itself is composed of a sequence of
elementary quantum gates \cite{BBC+:95}. So far, only little is known
about the implementation of POVMs even in quantum systems with a small
number of dimensions.  While some rather specific single-qubit
measurements have been studied \cite{Sasaki,Brandt,Decker}, not much
is known about the general problem of how to implement a POVM by a
unitary transform on the quantum register of a possibly larger space
followed by an orthogonal measurement in the computational basis.

When studying quantum circuits for families of POVMs questions about
the complexity of the required unitary transforms arise. In some cases
we can exploit the fact that they admit some additional symmetry. This
leads to the study of group-covariant POVMs which has been studied
extensively in the literature \cite{Davies,Sasaki,Ariano1,Ariano2}. As
a recent example we mention the construction of symmetric
informationally complete POVMs by means of suitable finite symmetry
groups \cite{RBSC:2004}.

The main contribution of this paper is a general method which computes an
embedding of group-covariant POVMs into orthogonal
measurements on a larger Hilbert space. A particular feature of the
computed embedding is that it uses the symmetry. This
in turn allows to apply known techniques for decomposing matrices with
symmetry to the unitary matrices obtained by this embedding. For
several cases this leads to families of {\em efficient} quantum
circuits implementing the given POVMs. 

\medskip
\textit{Outline.} In Section \ref{POVMdef} we briefly recall the
definition of POVMs. In Section \ref{POVMsym} we consider the
decomposition of matrices that have a symmetry with respect to a
group. This type of decomposition is a basic tool for our
constructions. We also define group-covariance of POVMs with respect
to a symmetry group and a group representation. Furthermore, we
explain how POVMs with this group-covariance are related to so-called
monomial representations of the symmetry group. In Section
\ref{POVMorth} we explain the general scheme for the construction of a
unitary transform that implements a group-covariant POVM. The basis
for this construction is the analysis of the intertwining space
between the group representation that is given by the group-covariance
of the POVM and the monomial representation. This is the starting
point for methods using fast quantum Fourier transforms as described
in Section \ref{POVMcovar}. Finally, in Section \ref{POVMexamples} we
give several examples of implementations of group-covariant POVMs.

\medskip
\textit{Notations.} We denote the field of complex numbers by
$\C$. The group of invertible $n\times n$ matrices is denoted by
$\GL_n(\C)$ and the subgroup consisting the unitary $n\times n$
matrices is denoted by ${\cal U}(n)$. We denote the identity matrix in
${\cal U}(n)$ by $\onemat_n$. If not denoted otherwise all matrices
are matrices over the complex numbers. The cyclic group of order $n$
is denoted by $\Z_n$. Representations are denoted by small Greek
letters, e.\,g., $\varphi$, $\psi$ etc. By abuse of notation we also
denote the trivial representation of degree $n$ (i.e. dimension $n$) 
by $\onemat_n$. The
base change of a matrix $A$ with respect to a matrix $B$ is denoted by
$A^B = B A B^\dagger$.  The direct sum of matrices and representations
is denoted by $A\oplus B$ and $\varphi \oplus \psi$ and the tensor
product is denoted by $A\otimes B$ and $\varphi \otimes \psi$,
respectively. We make frequent use of the Pauli matrices
\[
\sigma_x = \left(\begin{array}{rr}
0 & 1\\ 1 & 0 
\end{array}\right), \quad
\sigma_y = \left(\begin{array}{rr}
0 & -i\\ i & 0 
\end{array}\right), \quad
\sigma_z = \left(\begin{array}{rr}
1 & 0\\ 0 & -1 
\end{array}\right).
\]
A diagonal matrix with diagonal entries $\lambda_1, \ldots, \lambda_n$
is abbreviated by ${\rm diag}(\lambda_1, \ldots, \lambda_n)$.  We
denote the symmetric group on $n$ symbols by ${\cal S}_n$. To each
permutation $\sigma \in {\cal S}_n$ naturally corresponds the
permutation matrix $\sum_i \ket{\sigma(i)} \bra{i}$. By abuse of 
notation we identify $\sigma$ with the corresponding permutation
matrix. We often use the permutation matrix $S_m$ which corresponds to
the $m$-cycle $(1,2,\ldots, m)$ and the matrix $T_m = {\rm diag}(1,
\omega_m, \ldots \omega_m^{m-1})$ which contains the eigenvalues of
$S_m$. 
The basis states of an $n$-qubit system correspond to 
binary strings of length $n$.
Quantum circuits are written from the left to the right, and
the qubits are arranged such that the most significant qubit 
(characterizing the left-most symbol of a binary string)
is on
top.  Throughout the paper a matrix entry ``$\cdot$'' stands for zero.

%
%

\section{POVMs and orthogonal measurements}
\label{POVMdef}

A POVM for a quantum system with Hilbert space ${\mathbb C}^d$ is a
set $P=\{ A_1, \ldots, A_n \}\subseteq {\mathbb C}^{d \times d}$ of
non-negative operators, where $\sum_k A_k = \onemat_d$. For a more
general definition for POVMs with an infinite number of operators we
refer to \cite{Davies2}. For example, the set of matrices
\[
P_2=\left\{ \frac{1}{3} \left(
\begin{array}{rr} 1 & 1  \\ 1 &  1 \end{array} \right),  \ \frac{1}{3}
\left( \begin{array}{rr} 1 &  \omega  \\ \omega^2 & 1  \end{array}
\right), \ \frac{1}{3}   \left( \begin{array}{rr}  1  & \omega^2  \\
\omega & 1 \end{array} \right) \right\} \subseteq 
{\mathbb C}^{2\times 2},
\]
where $\omega = {\rm exp}(2\pi i /3)$ is a third root of unity,
defines a POVM on a system with corresponding Hilbert space ${\mathbb
C}^2$. Suppose that the state of the system is described by the
density matrix $\rho \in {\mathbb C}^{d \times d}$. Then for a general
POVM the probability $p_k$ for the result $k$ is given by $p_k={\rm
tr}(\rho A_k)$.  An {\em orthogonal} measurement is a POVM with
mutually orthogonal operators $A_k$, i.\,e., we have that $A_k A_l = A_l
A_k = 0$ for $k \not = l$.

In the following we restrict ourselves to rank-one operators
$A_k=\ket{\Psi_k}\bra{\Psi_k}$. Note that the POVM vectors
$\ket{\Psi_k}$ need not  be normalized and that the restriction to
operators of rank one is for some applications
justified by Davies' theorem \cite{Davies}.
It states that we can always find a POVM with rank-one operators that
maximizes the mutual information. The example $P_2$, which consists of
three rank-one operators, can be written as $P_2= \{
\ket{\Psi_1}\bra{\Psi_1},\ket{\Psi_2}\bra{\Psi_2},
\ket{\Psi_3}\bra{\Psi_3} \}$, where
\[
\ket{\Psi_1} = \frac{1}{\sqrt{3}}
\left(\begin{array}{c}1 \\ 1 \end{array} \right), \ \ket{\Psi_2}
=\frac{1}{\sqrt{3}} \left(\begin{array}{c}1 \\\omega^2
\end{array}\right),\quad {\rm and} \quad \ket{\Psi_3}   =
\frac{1}{\sqrt{3}}\left( \begin{array}{c}1 \\ \omega \end{array}
\right)
\] 
are the corresponding POVM vectors in ${\mathbb C}^2$. Neumark's
theorem \cite{Peres} states that it is possible to implement a POVM by
reducing it to an orthogonal measurement on a larger system.  We
briefly recall this construction. Let $P=\{ A_k \} = \{\ket{\Psi_k}
\bra{\Psi_k}\}$ be a POVM with $n$ operators that acts on the Hilbert
space ${\mathbb C}^d$. For $n > d$ the vectors $\ket{\Psi_k}$ cannot
be mutually orthogonal. Consequently, we have to extend the system by
at least $n-d$ dimensions in order to define an orthogonal measurement
with $n$ different measurement outcomes. We want to implement an
orthogonal measurement ${\tilde P} = \{ {\tilde A}_k \} =\{ \ket{
{\tilde \Psi}_k} \bra{ {\tilde \Psi}_k}\}$ on the system with $n$
dimensions such that ${\tilde P}$ corresponds to the POVM $P$ on the
subsystem with $d$ dimensions, i.\,e., $p_k = {\rm tr}(\rho A_k) =
{\rm tr}(\tilde{\rho} \tilde{A}_k)$. Here ${\tilde \rho} = \rho \oplus
0_{n-d}\in {\mathbb C}^{n \times n}$ where $0_{n-d}$ denotes the zero
matrix of size $n-d$ is the embedding of the state into the larger
system.

We write the POVM vectors $\ket{\Psi_k}$ as columns of the matrix $M=
\left( \ket{\Psi_1} \ldots \ket{\Psi_n} \right) \in {\mathbb C}^{d
\times n}$. In the following we refer to $M$ as the defining matrix
for the POVM $P$. Now, the operators ${\tilde A}_k = \ket{{\tilde
\Psi}_k} \bra{{\tilde \Psi}_k} \in {\mathbb C}^{n \times n}$ with
$\ket{{\tilde \Psi}_k} = \ket{\Psi_k} \oplus \ket{\Phi_k}$ are the
columns of the matrix
\[
{\tilde M} = \left( \begin{array}{ccc} \ket{\Psi_1} & \ldots
&\ket{\Psi_n} \\ \ket{\Phi_1} & \ldots & \ket{\Phi_n} \end{array}
\right) \in {\cal U}(n).
\]
Note that $\tilde{M}$ can be an arbitrary unitary matrix which
contains $M$ as upper part of size $d \times n$. Since $P$ is a POVM
we have $M M^\dagger = \sum_k \ket{\Psi_k} \bra{\Psi_k} = \sum_k A_k
=\onemat_d$, i.\,e., finding a suitable $\tilde{M}$ is always possible. For
example in case of $P_2$ we obtain the defining matrix
\[
M=
\frac{1}{\sqrt{3}} \left(
\begin{array}{ccc} 1 & 1 & 1 \\ 1 & \omega^2 & \omega \end{array}
\right) \in {\mathbb C}^{2 \times 3}
\] 
and one possible choice for $\tilde{M}$ is to add the row given by
$(1/\sqrt{3}) ( 1, \omega, \omega^2)$. Hence the rank-one
projectors corresponding to the orthogonal measurement ${\tilde M}$
are
\[
\ket{{\tilde \Psi}_1} =
\frac{1}{\sqrt{3}} \left( \begin{array}{c} 1 \\ 1 \\ 1 \end{array}
\right), \ \ket{{\tilde \Psi}_2} = \frac{1}{\sqrt{3}} \left(
\begin{array}{c} 1 \\ \omega^2 \\ \omega \end{array} \right), {\rm
and} \ \ \ket{{\tilde \Psi}_3} = \frac{1}{\sqrt{3}} \left(
\begin{array}{c} 1 \\ \omega \\ \omega^2 \end{array} \right).
\]
The probability distribution ${\tilde p}_k = {\rm tr}({\tilde \rho}
{\tilde A}_k)$ of the constructed orthogonal measurement equals the
distribution $p_k$ of the original POVM since
\[
{\tilde p}_k = {\rm
tr}\left({\tilde \rho} {\tilde A}_k
\right) = {\rm tr}\left(
\left(\rho \oplus 0_{n-d} \right) \left(\begin{array}{cc}
\ket{\Psi_k}\bra{\Psi_k} & \ket{\Psi_k} \bra{\Phi_k} \\ \ket{\Phi_k}
\bra{\Psi_k} & \ket{\Phi_k} \bra{\Phi_k}
\end{array} \right) \right) = {\rm tr}\left( \rho A_k \right) = p_k.
\]
The embedding into a larger system can be realized by
using an ancilla register of a quantum computer.
It consists of $l$ qubits 
such that $2^l\geq n-d$. They are initially in the state
$|0\dots 0\rangle$. Then the space $\C^d\otimes |0\dots 0\rangle$ 
is the subspace where the POVM acts on and $\C^d\otimes (\C^2)^{\otimes l}$ 
is the extension.
The density operator $\tilde{\rho}$ acts on an $n$ dimensional 
subspace of the joint system consisting of the original system
and the ancilla register.
 In the following we will assume that also the system
space $\C^d$ is embedded into the state space of some qubits.

As explained above,
we can implement the POVM with corresponding matrix $M$ by
applying the unitary transform ${\tilde M}^\dagger$ to the initial
state ${\tilde \rho}$ of the joint system followed by a measurement
in the computational basis. Note that for the special case where the
columns of $M$ are already orthogonal we have that ${\tilde M} =
M$. In this case by implementing the matrix $M^\dagger$
followed by a measurement in the computational basis we can 
perfectly distinguish between the columns of $M$. 

In principle, the construction of an appropriate matrix ${\tilde M}$
is simple since we just have to find mutually orthogonal rows that
lead to a unitary matrix. However, $k$  qubits  allow
POVMs with $n=2^k$ operators. Hence the size of $\tilde{M}$
is exponential in $k$.
The complexity to
implement a unitary matrix on $k$ qubits can be upper bounded by
$O(4^k)$ \cite{Varti}
and a generic element of ${\cal U}(2^k)$ will
indeed require an exponential number of elementary transforms
(e.g. one- and two-qubit-gates). Therefore we are interested in the
construction of a matrix ${\tilde M}$ that can be implemented
efficiently, if such a construction exists at all. While finding efficient
factorizations is a hard problem in general, the situation becomes
easier in some cases where we are given the additional structure of a
group-covariant POVM. In the following sections we will give a
definition of group-covariance and the related notion
of symmetry. Later, we exploit the symmetry of the matrix $M$
and give several examples of POVMs that have efficient quantum circuit
implementations.

%
%

\section{Group-covariant POVMs and matrices with symmetry}
\label{POVMsym}

In the following we give a precise mathematical definition of the
notion of {\em symmetry} of a matrix $M\in {\mathbb C}^{m \times
n}$. Later we define group-covariance of a POVM and show that the
group-covariance in a natural way leads to matrices with symmetry. For
the necessary background on finite groups and representations we refer
to standard textbooks such as \cite{HuppertI:83,Isaacs:76}.

We start with a finite group $G$ and a pair $(\varphi,\psi)$ of matrix
representations of $G$ which are compatible with the size of $M$,
i.\,e., $\varphi: G \to \GL_m(\C)$ and $\psi: G \to \GL_n(\C)$.
Following, \cite{Egner, Egner2} we call the triple $(G, \varphi,
\psi)$ a symmetry of $M$ if the identity $\varphi(g) M = M \psi(g)$
holds for all $g \in G$.  Sometimes we abbreviate this by using the
shorthand notation $\varphi M = M \psi$. Note that if $M$ is not a
square matrix the representations $\varphi$ and $\psi$ have different
degrees.

To give an example we let $\omega = \exp(2\pi i/3)$ and let $\alpha,
\beta,\gamma \in {\mathbb C}$. Then for all $j \in \{0,1,2\}$ we have
that
\[
\left(
\begin{array}{ccc}1 & \cdot &\cdot \\ \cdot &
\omega & \cdot \\ \cdot & \cdot &
\omega^2 \end{array} \right)^j \left( \begin{array}{ccc} 
\alpha & \alpha & \alpha \\ 
\beta & \beta \omega & \beta \omega^2 \\ \gamma & \gamma \omega^2 
& \gamma \omega \end{array} \right) = \left( 
\begin{array}{ccc} \alpha & \alpha & \alpha \\ 
\beta & \beta \omega & \beta \omega^2 \\ \gamma & \gamma \omega^2 
& \gamma \omega \end{array} \right) \left( \begin{array}{ccc} \cdot & \cdot
& 1 \\ 1 & \cdot & \cdot \\ \cdot & 1 & \cdot \end{array}\right)^j.
\]
Hence we obtain a symmetry which is given by the cyclic group
$\Z_3=\{0,1,2\}$ together with the two representations $\varphi,
\sigma: \Z_3 \to {\cal U}(3)$ given by $\varphi(1) = {\rm diag}(1,
\omega, \omega^2)$ and $\sigma(1) = (1,3,2)$.  

Note that given two representations $\varphi$, $\psi$ of a group $G$
the set of all matrices $M$ which fulfill $\varphi(g) M = M \psi(g)$
for all $g \in G$
is a vector space. It turns out that the matrices in this vector space
have a special form. Hence we explore its structure in more detail in
the following.

\begin{definition}[Intertwining space]
Let $G$ be a group and let $\varphi$, $\psi$ be representations of $G$
of degrees $n$ and $m$, respectively. Then 
\[
{\rm Int}(\varphi, \psi) := \{ M : \varphi(g) M = M \psi(g), \;
\mbox{for all} \; g \in G \}
\]
with $M\in \C^{n\times m}$
is called the intertwining space of $\varphi$ and $\psi$. 
\end{definition}

In the following we denote by $\varphi_1, \ldots, \varphi_k$ a list 
 of all pairwise inequivalent irreducible representations of
$G$. Recall that for any representation of a finite group it is always
possible to find a base change such that the corresponding
representation is a direct sum of irreducible representations
\cite{Isaacs:76}. For representations which are completely
decomposed into a direct sum of irreducibles the structure of the
intertwining space is known. This is
the content of the following theorem which follows directly
from Schur's Lemma (see \cite[Section \S 29]{CR:62}).
\begin{theorem}\label{intertwine}
Let $G$ be a finite group and $\varphi = \bigoplus_{i=1}^k
(\onemat_{n_i} \otimes \varphi_i)$ and $\psi = \bigoplus_{i=1}^k
(\onemat_{m_i} \otimes \varphi_i)$ two representations of $G$ which
have been completely decomposed into pairwise inequivalent
representations $\varphi_i, i=1,\ldots,k$. Then the intertwining space
of $\varphi$ and $\psi$ has the following structure:
\[
{\rm Int}(\varphi, \psi) = (\C^{n_1\times m_1} \otimes \onemat_{{\rm
deg}(\varphi_1)}) \oplus \ldots \oplus (\C^{n_k\times m_k} \otimes
\onemat_{{\rm deg}(\varphi_k)}).
\]
\end{theorem}

A matrix $A$ is called {\em block permuted} if there are permutation
matrices $P$ and $Q$ such that $P A Q = B_1 \oplus \ldots \oplus B_k$,
where $B_1, \ldots, B_k$ are (rectangular) matrices. For all $n,m,k
\in \N$ there exist permutation matrices $P_{n,m,k}$ and $Q_{n,m,k}$
such that for all $A \in \C^{n\times m}$ we have ${P_{n,m,k}} (A
\otimes \onemat_k) {Q_{n,m,k}} = \onemat_k \otimes A$. Hence we have
shown that the elements of the intertwining space of completely
reduced representations are block permuted.

We continue with an easy observation which turns out to be essential
for the approach of extending the symmetry of a given group-covariant
POVM to a measurement on a larger space. Suppose that $M\in {\rm
Int}(\varphi,\psi)$ and that the matrices $U$ and $W$ decompose the
representations $\varphi$ and $\psi$ into the direct sums, i.\,e., $U
\varphi U^\dagger = \varphi_1 \oplus \ldots \oplus \varphi_n$ and $V
\psi V^\dagger = \psi_1, \oplus \ldots \oplus \psi_m$. Then we can
rewrite $\varphi M = M \sigma$ as
\[
U^\dagger (\varphi_1 \oplus \ldots \oplus
\varphi_n) U M = M W^\dagger (\psi_1 \oplus \ldots \oplus \psi_m)
W.
\]
Multiplying this from the left by $U$ and from the right by
$W^\dagger$ shows that $C:=UMW^\dagger$ is an element of the
intertwining space ${\rm Int}(\varphi_1 \oplus \ldots \oplus
\varphi_n, \psi_1 \oplus \ldots \oplus \psi_m)$ of two completely
reduced representations. In particular, we can apply Theorem
\ref{intertwine} to determine the structure of $C$. In particular we
obtain that $C$ is block permuted and the size of the blocks depend on
the multiplicities and degrees of the irreducible representations
contained in $\varphi$ and $\psi$. 

Matrices with symmetry arise naturally in context of group-covariant
POVMs. We first give a definition of these POVMs and then establish a
connection between the notions of group-covariance and symmetry.

\begin{definition}[Group-covariant POVMs]
A POVM $P=\{ A_1, \ldots , A_n \}\subseteq {\mathbb C}^{d
\times d}$ with $A_k \not = A_l$ for $k\not = l$ is group-covariant
with respect to the group $G$ if there exists a projective unitary 
representation $\varphi: G \to {\cal U}(d)$ with $\varphi(g) \,
A_k \, \varphi(g)^\dagger \in P$ for all $g \in G$ and all $k$. 
\end{definition}
Note that a group-covariant POVM is also group-covariant for all
subgroups $H\leq G$ and the restriction of the representation
$\varphi$ to $H$. As a special case, the choice of the trivial
subgroup $H=\{1\}$ means that we do not use the symmetry of the POVM
at all.

A minor complication arises due to the fact that while the notion of
symmetry of matrices relies on ordinary, i.\,e., non-projective
representations, the definition of group-covariant POVMs relies on
projective representations. Therefore, we need a construction which
allows to transform the projective representation of the symmetry
group of a group-covariant POVM into a non-projective representation.
This connection is established using so-called {\em central
extensions} which is a method going back to I.~Schur. We briefly
recall this construction (see also \cite[Lemma
(11.16)]{Isaacs:76}). Let $\varphi: G \to \GL_d(\C)$ be a projective
representation of the group $G$. More precisely, we have $\varphi(gh)=
\gamma_{gh} \varphi(g)\varphi(h)$ for $g,h \in G$, where $\gamma_{gh}$
is a factor system. Let $H=\langle \gamma_{gh} : g,h \in G \rangle$ be
the group generated by the $\gamma_{gh}$. We consider the group
$\hat{G}$ consisting of the elements $(g,h)$ with $g \in G$ and $h\in
H$. The multiplication of two elements $(g,h)$ and
$(g^\prime,h^\prime)$ of $\hat{G}$ is defined by
$(g,h)(g^\prime,h^\prime)=(gg^\prime,\gamma_{gg^\prime} h
h^\prime)$. Then the map ${\tilde \varphi}((g, h))=h\varphi(g)$ is a
representation with ${\tilde \varphi}((g,1))=\varphi(g)$, i.\,e., the
representation ${\tilde \varphi}$ equals $\varphi$ on the elements
$(g,1)$ and the group $\hat{G}$ is a central extension of the
group $G$.

In the following we always assume $\varphi$ to be a non-projective
representation of the symmetry group $G$ by this construction. This is
justified since the set of POVM operators does not change by
switching from $G$ to a central extension $\hat{G}$ because scalar
multiples of the identity
operate trivial under conjugation. 

We now analyze the structure of the matrix $M$ corresponding to the
group-covariant POVM $P= \{ \ket{\Psi_k} \bra{\Psi_k} \}$ with
rank-one operators. Note that the phases of the vectors $\ket{\Psi_k}$
can be chosen arbitrarily without changing the POVM. Let $\varphi: G
\to {\cal U}(d)$ be the representation corresponding to
the symmetry of $P$. We then have the equation
\[
\varphi(g) \ket{\Psi_k} \bra{\Psi_k} \varphi(g)^\dagger =
\ket{\Psi_{\pi(g) k}} \bra{\Psi_{\pi(g)k}}
\] 
where $\pi: G \to S_n$ denotes a permutation representation of the
group $G$. Indeed, the equation
$\ket{\Psi_{\pi(g)j}}\bra{\Psi_{\pi(g)j}} =
\ket{\Psi_{\pi(g)k}}\bra{\Psi_{\pi(g)k}}$ implies
$\ket{\Psi_j}\bra{\Psi_j} = \ket{\Psi_k}\bra{\Psi_k}$ by conjugation
with $\varphi(g)^\dagger$ since $A_j \not = A_k$ for $j \not = k$.
Therefore, the map $\pi(g)$ is injective for all $g \in G$.  Since an
injective map on a finite set is also surjective the map $\pi(g)$
defines a permutation.

Next, we consider the action of $\varphi$ on the columns of the matrix
$M$.  As stated above the columns $\ket{\Psi_k}$ of $M$ can have
arbitrary phase factors. The action of $\varphi(g)$ on the columns of
$M$ can be described by the equation $\varphi(g)\ket{\Psi_k} = e^{i
\phi(g,k)} \ket{\Psi_{\pi(g)k}}$ where $\phi(g,k)$ depends on $k$, $g$
and the fixed phase factors of the vectors $\ket{\Psi_k}$. We identify
the columns $\ket{\Psi_k}$ with a basis $b_k$ of the vector space
${\mathbb C}^n$ in order to construct a representation that describes
the action of $\varphi$ on the columns of $M$.  With this
identification the action of $\varphi(g)$ corresponds to the map $b_k
\mapsto e^{i \phi(g,k)} b_{\pi(g)k}$.

By writing down the matrix corresponding to this map, we see that in
each row and each column there is precisely one entry different from
zero. Matrices having a structure like this are called {\em monomial
matrices}\footnote{Note that this terminology is somewhat unfortunate
since it has nothing to do with the monomials of which a polynomial is
comprised of. Still it is the standard terminology used in
representation theory.} \cite[Section \S 43]{CR:62}. Whenever the
images under a representation consist entirely of monomial matrices,
we denote this with an underscript, i.\,e., we write $\varphi_{\rm
mon}(g)$. Now, the two representations $\varphi$ and $\varphi_{\rm
mon}$ define the symmetry $\varphi M = M \varphi_{\rm mon}$ of the
matrix $M$.  The monomial representation $\varphi_{\rm mon}$ acts on
the columns of $M$. For each $g \in G$ it permutes the columns of $M$
and multiplies each column with a phase factor.

\begin{example}\rm
As an example in two dimensions we consider the following POVM:
\[
P= \left\{ \left( \begin{array}{cc} |\alpha|^2
& \alpha {\overline \beta} \\ {\overline \alpha} \beta & |\beta|^2
\end{array} \right), \left( \begin{array}{cc} |\alpha|^2 & -\alpha
{\overline \beta} \\ -{\overline \alpha} \beta & |\beta|^2 \end{array}
\right), \left( \begin{array}{cc} |\beta|^2 & {\overline \alpha} \beta
\\ \alpha {\overline \beta} & |\alpha|^2 \end{array} \right), \left(
\begin{array}{cc} |\beta|^2 & -{\overline \alpha} \beta \\ -\alpha
{\overline \beta} & |\alpha|^2 \end{array} \right) \right\} \subseteq
{\mathbb C}^{2 \times 2}
\]
with $\alpha, \beta \in {\mathbb C}$ and $|\alpha|^2 + |\beta|^2 =
1/2$. Then $P$ is covariant with respect to $\Z_2 \times
\Z_2$. The corresponding projective representation $\varphi: \Z_2
\times \Z_2 \to {\cal U}(2)$ is defined by the equations
\[
\varphi(0,0) = \onemat_2, \;
\varphi(0,1) = \sigma_z, \;
\varphi(1,0) = \sigma_x, \;
\varphi(1,1) = \sigma_z \sigma_x
\]
where $(0,0)$, $(0,1)$, $(1,0)$ and $(1,1)$ denote the elements of the
group $\Z_2 \times \Z_2$.

For this projective representation of $\Z_2 \times \Z_2$ a simple
computation shows that the central extension $\hat{G}$ of $\Z_2 \times
\Z_2$ is isomorphic to the dihedral group with eight elements. In the
following it is sufficient to consider the definition of the
representation on the elements $((0,1),1)$ and $((1,0),1)$ since these
elements generate $\hat{G}=\{ (g,h): g \in \Z_2 \times \Z_2 , h \in
\{\pm 1\} \}$. We can choose
\[
M =
\left( \begin{array}{rrrr} \alpha & \alpha & \beta & \beta \\
\beta & -\beta & \alpha & -\alpha \end{array} \right)  
\in {\mathbb C}^{2 \times 4}
\]
or a matrix with the same columns (up to an arbitrary phase factor for 
each column).  This leads to a symmetry group given by the monomial
representation
\[
\varphi_{\rm mon}((0,1),1)=\left(
\begin{array}{rrrr}\cdot&1&\cdot&\cdot\\1&\cdot&\cdot&\cdot\\\cdot&\cdot&\cdot&1\\\cdot&\cdot&1&\cdot\end{array}\right)
\quad {\rm and} \quad \varphi_{\rm mon}((1,0),1)= 
\left( \begin{array}{rrrr}\cdot&\cdot&1&\cdot\\\cdot&\cdot&\cdot&-1\\1&\cdot&\cdot&\cdot\\\cdot&-1&\cdot&\cdot\end{array}\right).
\]
For a different choice of phase factors we obtain another representation
 $\varphi_{\rm mon}$. The modified pair of representations $\varphi, 
\varphi_{\rm mon}$  also defines a symmetry of $M$.
\end{example}

An important special case of group-covariant POVMs are {\em
group-generated} POVMs which we describe next. Let $G$ be a group and
$\varphi: G \to {\mathbb C}^{d\times d}$ an (ordinary) unitary
representation. A group-generated POVM is described by the POVM
vectors $\varphi(g)\ket{\Psi}$ for $g \in G$ and an initial vector
$\ket{\Psi} \in {\mathbb C}^d$.  The corresponding operators of the
POVM are given by $A_g =\varphi(g)\ket{\Psi}\bra{\Psi}
\varphi(g)^\dagger$ for $g \in G$.  In other words, all POVM vectors
are obtained by the initial vector $\ket{\Psi}$ under the operation of
the group $G$, i.\,e., they form an orbit. Obviously, a
group-generated POVM is a group-covariant POVM with a single orbit
under the action of the group.  With this construction, the phase
factors of the POVM vectors $\varphi(g)\ket{\Psi}$ are fixed by the
phase factor of the initial vector $\ket{\Psi}$. The phase factors
$e^{i \phi(g,k)}$ of the monomial representation $\varphi_{\rm mon}$
corresponding to $\varphi$ equal $1$. As a consequence, the monomial
representation $\varphi_{\rm mon}$ equals the regular representation
of $G$ where we have to consider a fixed order of the elements of $G$.

Note that the operators $\{ \varphi(g) \ket{\Psi} \bra{\Psi}
\varphi(g)^\dagger \}$ in general do not define a POVM for arbitrary
representations $\varphi$ and initial vectors $\ket{\Psi}$. However,
if $\varphi$ acts irreducibly one has (after appropriate normalization)
for every 
vector $\ket{\Psi}$ the equation $ \sum_{g \in G} \varphi(g)
\ket{\Psi}\bra{\Psi} \varphi(g)^\dagger = \onemat_d$.

%
%

\section{Construction of the orthogonal measurement}
\label{POVMorth}

Following the previous section we can arrange the vectors which
correspond to the elements of a POVM with rank one projectors into the
columns of a matrix $M$. We have seen that in case of a
group-covariant POVM the matrix $M \in {\mathbb C}^{d \times n}$
always has the symmetry $\varphi M = M \varphi_{\rm mon}$ where
$\varphi$ is the given representation and $\varphi_{\rm mon}$ is a
monomial representation. Both representations are representations of
the symmetry group of the group-covariant POVM. 
We know that both representations are
equivalent to direct sums of irreducible representations.  Hence we
can find unitary matrices $U$ and $W$ such that $U\varphi U^\dagger =
\varphi_1 \oplus \ldots \oplus \varphi_n$ and $W\varphi_{\rm
mon}W^\dagger = \sigma_1 \oplus \ldots \oplus \sigma_m$ where the
$\varphi_k$ and the $\sigma_l$ denote irreducible representations of
the group $G$. In general, we can write the equation $\varphi M = M
\varphi_{\rm mon}$ as
\[
U^\dagger (\varphi_1 \oplus \ldots \oplus \varphi_n) U M = M W^\dagger
(\sigma_1 \oplus \ldots \oplus \sigma_m) W.
\]
This is equivalent to $C=UMW^\dagger \in T := {\rm Int}( \varphi_1
\oplus \ldots \oplus \varphi_n,\sigma_1 \oplus \ldots \oplus
\sigma_m)$. Conversely, a matrix $C$ which is contained in this
intertwining space and has orthogonal rows defines (up to an
appropriate normalization) a group-covariant POVM with corresponding
matrix $M= U^\dagger C W$.

For a given matrix $M\in {\mathbb C}^{d \times n}$ we now consider the
construction of a unitary matrix ${\tilde M} \in {\cal U}(n)$ such
that ${\tilde M}$ contains $M$ as upper part, i.\,e., we are looking
for a matrix ${\tilde M}$ such that 
\[
{\tilde M} = \left(
\begin{array}{c} M \\ \hline N \end{array} \right),
\]
where $N \in \C^{(n-d)\times n}$. In addition to this we intend to get
the symmetry $(\varphi \oplus \varphi^\prime ) {\tilde M} = {\tilde M}
\varphi_{\rm mon}$ with an appropriate representation $\varphi^\prime:
G \to {\cal U}(n-d)$. If we succeed in constructing
an appropriate representation $\varphi^\prime$ and matrix ${\tilde M}$
then we have the equation $\varphi \oplus \varphi^\prime = {\tilde M}
\varphi_{\rm mon} {\tilde M}^\dagger$, i.\,e., the representation
$\varphi \oplus \varphi^\prime$ has to be equivalent to $\varphi_{\rm
mon}$. In other words, each irreducible representation of $G$ is
contained the same number of times in $\varphi \oplus \varphi^\prime$
and in $\varphi_{\rm mon}$. Furthermore, from the decompositions
$(U\oplus \onemat_{n-d})(\varphi \oplus \varphi^\prime) (U^\dagger \oplus
\onemat_{n-d})=\sigma_{\tau(1)} \oplus \ldots \oplus \sigma_{\tau(m)}$ and
$W\varphi_{\rm mon}W^\dagger=\sigma_1 \oplus \ldots \oplus\sigma_m$ we
obtain that 
\begin{equation}\label{Gleichung} 
(U \oplus \onemat){\tilde M} W^\dagger \in {\tilde T}:={\rm
Int}(\sigma_{\tau(1)} \oplus \ldots \oplus \sigma_{\tau(m)}, \sigma_1
\oplus \ldots \oplus \sigma_m ) \subseteq {\mathbb C}^{n \times n}.
\end{equation} 
The permutation $\tau$ used in eq.~(\ref{Gleichung}) is a suitable
reordering of the irreducible representations. The structure of the
intertwining space ${\tilde T}$ is known from Theorem \ref{intertwine}
since we can compute the irreducible representations $\sigma_j$ from
$\varphi_{\rm mon}$.

In the following discussion we consider the construction of
$\varphi^\prime$ and ${\tilde M}$. Our goal is to show that the
construction of $\varphi^\prime$ that makes $U \varphi U^\dagger
\oplus \varphi^\prime$ equal to $W\varphi_{\rm mon}W^\dagger$ up to a
permutation $\tau$ of the irreducible components is always possible.

Important for the extension of $M$ to $\tilde{M}$ will be the
following theorem which characterizes the relations of two
representations in case there is an intertwiner of maximal possible
rank.  Recall that $\psi_1$ is a constituent of $\psi_2$ if and only
if there is a base change $U$ such that $U^{-1} \psi_2(g) U =
\psi_1(g) \oplus \psi_1^\prime(g)$ where $\psi_1^\prime$ is a
representation of $G$.

\begin{theorem}\label{interRank}
Let $G$ be a finite group and let $\psi_1, \psi_2$ be representations
of $G$ of degrees $d_1 = {\rm deg}(\psi_1)$ and $d_2 = {\rm
deg}(\psi_2)$, respectively. Let $M \in \C^{d_1 \times d_2}$ be a
matrix with $\psi_1(g) M = M \psi_2(g)$ for all $g \in G$ and ${\rm
rk}(M) = {\rm deg}(\psi_1)$. Then $\psi_1$ is a constituent of
$\psi_2$.
\end{theorem}

\noindent
{\bf Proof:} Let $M$ be such that $\psi_1(g) M = M \psi_2(g)$ and let
$\varphi_1, \ldots, \varphi_k$ be a complete set of pairwise
inequivalent irreducible representations of $G$. Since $\psi_1$,
$\psi_2$ are representations of a finite group over the field
of complex numbers
we find unitary matrices $U, W$ such that $U \psi_1 U^\dagger =
\bigoplus_{i=1}^k m_i \varphi_i$ and $W \psi_2 W^\dagger =
\bigoplus_{i=1}^k n_i \varphi_i$, where the multiplicities $m_i$ and
$n_i$ are non-negative integers. We have to show that actually $m_i
\leq n_i$ for all $i=1, \ldots, k$.

From $\psi_1 M = M \psi_2$ and by the choice of $U$ and $W$ we obtain
that $(\bigoplus m_i \varphi_i) (U M W^\dagger) = (U M W^\dagger)
(\bigoplus n_i \varphi_i)$, i.\,e., we have that $UMW^\dagger \in {\rm
Int}(\bigoplus_{i=1}^k m_i \varphi_i, \bigoplus_{i=1}^k n_i
\varphi_i)$.  By the remarks following Theorem \ref{intertwine} we
know that there are permutation matrices $P$ and $Q$ such that $M_0 :=
P (U M W^\dagger) Q = (\onemat_{{\rm deg}(\varphi_1)} \otimes B_1) \oplus
\ldots \oplus (\onemat_{{\rm deg}(\varphi_k)} \otimes B_k)$ where each
$B_i \in \C^{m_i \times n_i}$. Multiplication with invertible
matrices preserves the property that $M$ and hence also
$M_0$ have full rank (given by ${\rm deg}(\psi_1)$). On the other hand
we know that the rank of a block diagonal matrix is given by the sum
of the ranks of the blocks. Hence ${\rm rk}(M_0) = \sum_{i=1}^k {\rm
deg}(\varphi_i) \cdot {\rm rk}(B_i)$ which shows that each $B_i$ must
have full rank. Since $B_i$ is an $m_i \times n_i$ matrix this in
particular implies that $m_i \leq n_i$. This shows that $\psi_1$ is a
constituent of $\psi_2$. \qed

We now use Equation (\ref{Gleichung}) to construct the matrix ${\tilde
M}$ for the implementation of a group-covariant POVM. Having
determined $U$ and $W$ we can compute the matrix $C=UMW^\dagger \in
{\rm Int}(U \varphi U^\dagger, W \varphi_{\rm mon} W^\dagger)$. The
number of times each irreducible representation has to occur in
$\varphi^\prime$ can be computed. Since the structure of the
intertwining space ${\tilde T}= {\rm Int}(U\varphi U^\dagger \oplus
\varphi^\prime, W\varphi_{\rm mon}W^\dagger)$ is known we can extend
$C$ to an arbitrary unitary matrix ${\tilde C}$ of the intertwining
space ${\tilde T}$. This extension is always possible since both
representations $U\varphi U^\dagger \oplus \varphi^\prime$ and
$W\varphi_{\rm mon}W^\dagger$ contain each irreducible representation
the same number of times. The matrix $C$ defines some of the rows of
$A$. Since $M$ defines a POVM the rows are mutually
orthogonal. Consequently, the matrix components of ${\tilde C}$
corresponding to an irreducible representation can be chosen under the
constraint that they are orthogonal. We now have that for any $V \in
{\cal U}(n-d)$ the matrix ${\tilde M} = (U^\dagger \oplus V^\dagger)
{\tilde C} W$ yields a unitary that extends the matrix
$M$ and has the symmetry we wanted to construct.

Hence, we obtain the following algorithm to construct an orthogonal
measurement which realizes the given POVM and preserves the symmetry.

\begin{algorithm}\label{POVMalg} \rm
Let $P=\{A_1, \ldots, A_n\} \subseteq \C^{d\times d}$ be a POVM. Then
the following steps implement $P$ by a von Neumann measurement on a
larger space. 
\begin{enumerate}
\item Write the rank-one operators $A_k=\ket{\Psi_k}\bra{\Psi_k}$ 
of the POVM as columns of the matrix $M\in {\mathbb C}^{d \times n}$.

\item Determine an appropriate symmetry group with corresponding
representation $\varphi: G \to {\cal U}(d)$.

\item Compute the monomial representation 
$\varphi_{\rm mon}: G \to {\cal U}(n)$.

\item Find a matrix $U\in {\cal U}(d)$
that decomposes $\varphi$ into irreducible representations where
equivalent ones are equal. 

\item Find a matrix $W \in {\cal U} (n)$ 
that decomposes $\varphi_{\rm mon}$ into irreducible representations
where equivalent ones are equal. 

\item Construct the representation $\varphi^\prime$ such that
$U\varphi U^\dagger \oplus \varphi^\prime$ is equal to $W\varphi_{\rm
mon}W^\dagger$ up to a permutation $\tau$ of the irreducibles.

\item Construct ${\tilde C} \in {\cal U}(n)$ 
that contains $C= UMW^\dagger \in {\mathbb C}^{d \times n}$ as upper
part and is in the intertwining space ${\tilde T}$ of 
$U\varphi U^\dagger \oplus
\varphi^\prime$ and $W\varphi_{\rm mon}W^\dagger$.

\item Choose an arbitrary unitary matrix 
$V\in {\cal U}(n-d)$.

\item Compute ${\tilde M}=(U^\dagger \oplus V^\dagger)
{\tilde C}W \in {\cal U}(n)$.
\end{enumerate}
Then ${\tilde M}^\dagger$ implements the POVM $P$ by a von Neumann
measurement on a larger space, i.\,e., for any state $\rho$ on the
original $d$-dimensional system we have that $p_k = {\rm
tr}({\tilde \rho}{\tilde A}_k) = \langle {\tilde \Psi}_k | {\tilde
\rho} | {\tilde \Psi}_k \rangle$. Here $\ket{{\tilde \Psi}_k}$ denote
the rows of ${\tilde M}$ and ${\tilde \rho} = \rho \oplus 0_{n-d}$ is
the embedding of $\rho$ to a state of an $n$-dimensional system.
\end{algorithm}

\begin{example}\rm
We consider the example of the previous section with the matrix
\[
M=\left( \begin{array}{rrrr}\alpha & \alpha & \beta &
\beta \\ \beta & -\beta & \alpha &-\alpha \end{array}\right) \in
{\mathbb C}^{2 \times 4}
\]
and the group $G=\{ (g,h): g \in \Z_2 \times \Z_2, h \in \{\pm 1\} \}$
which is isomorphic to the dihedral group of order eight.  The
representation $\varphi: G \to {\cal U}(2)$ is given by
$\varphi((0,1),1) = \sigma_z$ and $\varphi((1,0),1) = \sigma_x$. We
have $U=\onemat_2$ and $U\varphi U^\dagger = \varphi$ since the
representation $\varphi$ is already irreducible. An elementary
computation shows that the corresponding monomial representation
$\varphi_{\rm mon}$ is given by
\[
W \varphi_{\rm mon}((0,1),1) W^\dagger = \left(
\begin{array}{rrrr}1&\cdot&\cdot&\cdot\\\cdot&-1&\cdot&\cdot\\ 
\cdot&\cdot&1&\cdot\\\cdot&\cdot&\cdot&-1\end{array}\right)\ {\rm and} \ W \varphi_{\rm
mon}((1,0),1) W^\dagger = \left( \begin{array}{rrrr}\cdot&1&\cdot&\cdot\\1&\cdot&\cdot&\cdot\\
\cdot&\cdot&\cdot&1\\\cdot&\cdot&1&\cdot\end{array}\right)
\]
with the unitary matrix
\[
W = \frac{1}{\sqrt{2}} \left( \begin{array}{rrrr}1&1&\cdot&\cdot \\ \cdot&\cdot&1&-1
\\ \cdot&\cdot&1&1 \\ 1&-1&\cdot&\cdot \end{array} \right) \in
{\cal U}(4).
\]
Therefore, $\varphi_{\rm mon}$ contains the irreducible representation
$\varphi$ twice, i.\,e., $W\varphi_{\rm mon} W^\dagger =
\varphi \oplus \varphi$. 

With the matrices $M \in {\mathbb C}^{2\times 4}, U\in {\cal U}(2),$ 
and $W \in {\cal U}(4)$ as above we find that
$ C=UMW^\dagger = \sqrt{2} \left(\begin{array}{cc}\alpha &
\beta \end{array} \right) \otimes \onemat_2 \in {\mathbb C}^{2 \times
4}, $ which is an element of the intertwining space 
\[
{\rm
Int}(\varphi, W\varphi_{\rm mon}W^\dagger)= {\rm Int}(\varphi,\varphi
\oplus \varphi)\,.
\]
Since we have $W \varphi_{\rm mon} W^\dagger = \varphi \oplus
\varphi$, we have to choose $\varphi^\prime = \varphi$. The
intertwining space ${\tilde T}$ is given by
\[
{\tilde T}={\rm
Int}(\varphi \oplus \varphi, \varphi\oplus \varphi) = \left\{ \left(
\begin{array}{cccc} \lambda_{11}&\cdot&\lambda_{12}&\cdot\\
\cdot&\lambda_{11}&\cdot&\lambda_{12}\\
\lambda_{21}&\cdot&\lambda_{22}&\cdot \\
\cdot&\lambda_{21}&\cdot&\lambda_{22}\end{array} \right) \; : \;
\lambda_{ij} \in {\mathbb C} \right\} \subseteq {\mathbb C}^{4 \times
4}.
\]
In our example, the matrix $C=UMW^\dagger$ defines the first two rows
of the matrix $\tilde{C} \in {\tilde T}={\rm Int}(\varphi \oplus
\varphi, \varphi \oplus \varphi)$.

In particular, we have the equations $\lambda_{11}=\sqrt{2} \alpha$
and $\lambda_{12} = \sqrt{2} \beta$. For example, it is possible to
choose $\lambda_{21} = \sqrt{2} \,{\overline \beta}$ and
$\lambda_{22}=- \sqrt{2} \, {\overline \alpha}$ for $\alpha, \beta \in
{\mathbb C}$ to obtain the unitary matrix
\[
{\tilde C} = \sqrt{2}
\left(
\begin{array}{rrrr} \alpha & \cdot & \beta & \cdot \\ 
\cdot & \alpha & \cdot & \beta \\ {\overline \beta} & \cdot & -
{\overline \alpha} & \cdot \\ \cdot & {\overline \beta} & \cdot &
- {\overline \alpha} \end{array} \right)\in {\cal U}(4)
\]
which has the symmetry $(\varphi \oplus \varphi) {\tilde C} = {\tilde
C} (\varphi \oplus \varphi)$. With ${\tilde M} = (U^\dagger \oplus
V^\dagger) {\tilde C} W$ and $V=\onemat_2$ we compute the matrix
\[
{\tilde M}= \left( \begin{array}{rrrr} \alpha & \alpha & \beta &
\beta \\ \beta & -\beta & \alpha & -\alpha \\ {\overline \beta} &
{\overline \beta} & - {\overline \alpha} & - {\overline \alpha} \\ -
{\overline \alpha} & {\overline \alpha} & {\overline \beta} & -
{\overline \beta} \end{array}\right) \in {\cal U}(4) 
\]
that contains $M$ as upper part and has the symmetry $(\varphi \oplus
\varphi) {\tilde M} = {\tilde M} \varphi_{\rm mon}$.  Note that all 
unitary matrices $V\in {\cal U}(2)$ give rise to possible extensions
${\tilde M}$. 
\end{example}

%
%

\section{Efficient implementations of group-covariant POVMs}
\label{POVMcovar}

From the general construction of a von Neumann measurement which
realizes a given POVM using the symmetry of the
POVM we now turn to the question of decomposing the unitary 
$\tilde{M}$ into gates. This can be seen as a first step towards the more
general question of how POVMs can
be implemented efficiently on a quantum computer.

When speaking about the efficiency, we mean the cost of implementing
the POVM as a von Neumann measurement on a larger Hilbert space,
i.\,e., the number of elementary gates we need to actually implement
the necessary unitary operation on this bigger space.  First note that
the discussed construction of ${\tilde M}$ has several degrees of
freedom:
\begin{enumerate}
\item[$\bullet$] The matrix ${\tilde C}$ that contains $C$ as upper 
part can be chosen arbitrarily. The matrix ${\tilde C}$ has to be a
unitary matrix in the intertwining space ${\tilde T}$.

\item[$\bullet$]The matrix $V\in {\cal U}(n-d)$ 
can be an arbitrary unitary matrix.

\item[$\bullet$] The order and phase factors of the POVM vectors in the matrix
$M$ can be chosen arbitrarily. However, it must be possible to deduce
the applied POVM operator from the result of the orthogonal
measurement efficiently.

\item[$\bullet$] The permutation $\tau$ of the irreducible representations
in $U \varphi U^\dagger \oplus \varphi^\prime$ can be chosen arbitrarily.

\item[$\bullet$] The symmetry group $G$ can be restricted to subgroups
$H \leq G$ which might lead to different realizations of the POVM. 

\end{enumerate}

The constructions depend on the symmetry group $G$ we consider for the
POVM. Sometimes, we can obtain simple implementations by restricting
the symmetry group to a subgroup $H \leq G$.  If we consider a
subgroup $H$ of $G$ and construct the POVM with respect to $H$ we have
several changes in the construction compared to the construction with
the group $G$. On the one hand, the number of occurrences of the
irreducible representations in $\varphi_{\rm mon}$ increase. On the
other hand the number of inequivalent irreducible representations of
the symmetry group decreases. Consequently, the matrices of the
intertwining spaces are more complex since there are more irreducible
representations in $\varphi$ and $\varphi_{\rm mon}$ that are
equivalent. As a tradeoff we have that the complexity of the transform
$W$ decreases. The circuits constructed in \cite{Decker} show that the
restriction of the symmetry group to a cyclic subgroup can lead to
efficient algorithms in some cases.

Let $G$ be a finite group and $\{\varphi_1, \ldots, \varphi_k\}$ a
system of representatives for the irreducible representations of
$G$. Let the coefficients of these representation be indexed by the
list $L^\prime := [(m; i, j), 1\leq m \leq k, 1\leq i, j \leq {\rm
deg}(\varphi_m)]$.  Furthermore, let the elements of $G$ be indexed by
the list $L$.  Then the matrix $ 1/\sqrt{|G|}(\sqrt{{\rm
deg}(\varphi_m)} \, \varphi_m(g)_{ij})_{(m;i, j), g}$ is unitary and
is called a Fourier transform (or DFT for short) for $G$
\cite{Beth:87,CB:93} (with respect to $L$ and $L^\prime$).

For several groups it is known how to realize a DFT efficiently on a
quantum computer \cite{Beals:97,PRB:99,MRR:2004}.  In these cases the
symmetry $\varphi_{\rm mon}$ can be decomposed efficiently whenever we
have that (i) $\varphi_{\rm mon}$ is a regular
representation of $G$ and that (ii) the DFT for $G$ can be computed
efficiently.  Note that the computational complexity of this von
Neumann measurement depends essentially on the complexity of
implementing $\DFT_G$ in terms of elementary quantum gates. Hence we
obtain several families of POVMs for which the monomial representation
$\varphi_{\rm mon}$ can be decomposed efficiently. The complexity of
the corresponding POVM then depends on the remaining matrices $C$,
$U$, and $W$ used in Algorithm \ref{POVMalg}.

%
%

\section{Examples}
\label{POVMexamples}

In this section we apply the methods discussed in the preceding
sections to some examples of group-covariant POVMs. We exploit the
symmetry of group-covariant POVMs with respect to cyclic groups,
dihedral groups, and Weyl-Heisenberg groups in order to construct
quantum circuits for the implementation of these POVMs. Quantum
circuits for the implementation of group-covariant POVMs on a single
qubit with respect to the cyclic and dihedral groups are also
discussed in \cite{Decker}.  

\subsection{Cyclic groups}\label{zyklische Gruppen} 

Let $\Z_n=\{0,1, \ldots, n-1\}$ be a cyclic group with $n$ elements
and let $\omega={\rm exp}(2\pi i/n)$ be a primitive $n$th root of
unity. On a $d$-dimensional Hilbert space we consider a
group-generated POVM with respect to the representation $\varphi: \Z_n
\to {\cal U}(d)$ that is defined on the generator by
$\varphi(1) = {\rm diag}( 1, \omega,\omega^2,\ldots, \omega^{d-1})$. With an appropriate initial vector
$\ket{\Psi}\in {\mathbb C}^d$ the elements $\varphi(g)\ket{\Psi}$ for
$g \in \Z_n$ define a POVM. In the following, we only consider the
vector $\ket{\Psi}=1/\sqrt{n}(1, \ldots ,1)^T\in {\mathbb C}^d$. This
vector leads to the POVM with the defining matrix
\begin{equation}\label{x1}
M=\frac{1}{\sqrt{n}}\left( \begin{array}{ccccc} 1 & 1 & 1&\ldots & 1\\
1 & \omega & \omega^2 & \ldots & \omega^{n-1}\\
\vdots & \vdots & \vdots        & \ddots & \vdots\\
1    & \omega^{d-1} & \omega^{2(d-1)}&\ldots & \omega^{(n-1)(d-1)}
\end{array}\right) \in {\mathbb C}^{d \times n}.
\end{equation} 

The matrix $M\in {\mathbb C}^{d \times n}$ has the symmetry $\varphi M
= M \varphi_{\rm mon}$ where $\varphi_{\rm mon}(1) = (1, 2, \ldots,
n)$.  The representation $\varphi_{\rm mon}$ is the regular
representation of the cyclic group where the elements are ordered as
$[0,1, \ldots, (n-1)]$. With the Fourier matrix
\[
F_n=\frac{1}{\sqrt{n}}\left( \omega^{jk}
\right)_{j,k=0}^{n-1} \in {\cal U}(n)
\] 
we can write $F_n\, \varphi_{\rm mon}(1) \, F_n^\dagger={\rm
diag}(1,\omega,\omega^2, \ldots,\omega^{n-1})$. This shows that the
Fourier transform decomposes the regular representation of $\Z_n$ into
a direct sum of irreducible representations.

According to the preceding discussion (and notation) we have that
$U=\onemat_d$ and $W=F_n$. As a consequence we have the equation
$C=UMW^\dagger = MF_n^\dagger$. More precisely, we have
$C=MF_n^\dagger = {\rm diag}(1,1,\ldots, 1) \in {\mathbb C}^{d \times
n}$.

We now consider the construction of the matrices ${\tilde C}$ and
${\tilde M}$. The representation $\varphi: \Z_n \to {\cal U}(d)$ 
with $\varphi(1)={\rm diag}(1,\omega, \omega^2,\ldots,
\omega^{d-1})$ contains the irreducible representations $1 \mapsto
(\omega^k)$ for all $k \in \{0,\ldots d-1\}$. The representation
$F_n\varphi_{\rm mon}F_n^\dagger: \Z_n \to {\cal U}(n)$
with $F_n\varphi_{\rm mon}(1)F_n^\dagger = {\rm
diag}(1,\omega,\omega^2, \ldots,\omega^{n-1})$ contains the
irreducible representations $1 \mapsto (\omega^k)$ for all $k \in
\{0,1,\ldots, n-1\}$. Following Algorithm \ref{POVMalg} from Section
\ref{POVMorth}, we choose $\varphi^\prime$ with $\varphi^\prime(1) =
{\rm diag}(\omega^d, \ldots, \omega^{n-1})$ in order to obtain
$\varphi \oplus \varphi^\prime = F_n \varphi_{\rm mon}
F_n^\dagger$. Since each irreducible representation $1 \mapsto
(\omega^k)$ with $k \in \{0,1,\ldots, n-1\}$ has dimension one and the
irreducible representations defined by $1 \mapsto (\omega^k)$ are
inequivalent for different $k$ we have the intertwining space
\[
{\tilde T} = {\rm Int}(\varphi\oplus \varphi^\prime, F_n \varphi_{\rm mon}
F_n^\dagger) = \{ {\rm diag}(\lambda_1, \ldots,\lambda_n) : \lambda_j
\in {\mathbb C} \} \subseteq {\mathbb C}^{n \times n}.
\]
We have to find a  matrix ${\tilde C} \in {\cal U}(n)$ 
in the intertwining space ${\tilde T}$ that has the matrix $C \in
{\mathbb C}^{d \times n}$ as upper part. As stated above, the matrix
$M \in {\mathbb C}^{d \times n}$ defines $\lambda_j = 1$ for $j \in
\{0,1, \ldots, d-1\}$. Since ${\tilde C}$ has to be a unitary matrix
we have to choose $\lambda_j$ with the absolute value $|\lambda_j|=1$
for $j \in \{d, \ldots, n-1\}$.

In order to simplify the matrices we set $\lambda_j=1$ for all $j \in
\{d, \ldots , n-1\}$.  With these elements $\lambda_j$ we have the
equation ${\tilde C}=\onemat_n$. Furthermore, we choose
$V=\onemat_{n-d}$ in Algorithm \ref{POVMalg} from Section \ref{POVMorth}
leading to $U\oplus V=\onemat_n$.  Consequently, we obtain the
equation
\[
{\tilde M}^\dagger= W^\dagger {\tilde
C}^\dagger (U \oplus V)=F_n^\dagger \onemat_n \onemat_n =
F_n^\dagger.
\] 
This equation shows that the inverse Fourier transform ${\tilde
M}^\dagger =F_n^\dagger$ is a unitary transform that implements the
group-covariant POVM with defining matrix (\ref{x1}). Recall that
for $n=2^k$ where $k\in \N$ the Fourier transform can be implemented
efficiently on a qubit register \cite{Coppersmith:94,Nielsen}.

\subsection{Dihedral groups}

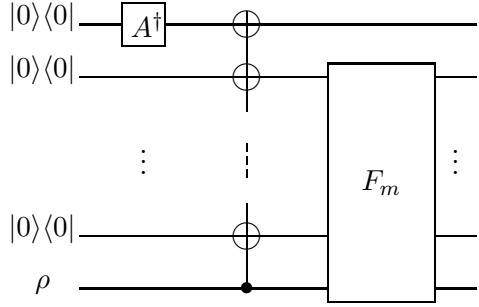
\begin{figure}
\centerline{\unitlength1pt
\begin{picture}(16,100)
\multiput(0,0)(0,20){2}{\line(1,0){16}}
\multiput(0,80)(0,20){2}{\line(1,0){16}}
\end{picture}%
\begin{picture}(16,100)
\put(0,92){\framebox(16,16){$\;A^{\dagger}$}}
\multiput(0,80)(0,20){1}{\line(1,0){16}}
\multiput(8,44)(0,4){3}{\makebox(0,0){.}}
\multiput(0,0)(0,20){2}{\line(1,0){16}}
\end{picture}%
\begin{picture}(16,100)
\multiput(0,0)(0,20){2}{\line(1,0){16}}
\multiput(0,80)(0,20){2}{\line(1,0){16}}
\end{picture}%
\begin{picture}(30,100)
\multiput(0,0)(0,20){2}{\line(1,0){30}}
\multiput(0,80)(0,20){2}{\line(1,0){30}}
\multiput(15,20)(0,20){1}{\circle{10}}
\multiput(15,80)(0,20){2}{\circle{10}}
\multiput(15,42)(0,5){3}{\line(0,1){3}}
\put(-65,0){$\rho$} 
\put(-75,20){$|0\rangle \langle 0|$} 
\put(-75,100){$|0\rangle \langle 0|$} 
\put(-75,80){$|0\rangle \langle 0|$} 
\put(15,67){\line(0,1){38}}
\put(15,0){\makebox(0,0){$\bullet$}}
\put(15,0){\line(0,1){32}}
\end{picture}%
\begin{picture}(16,100)
\multiput(0,0)(0,20){2}{\line(1,0){16}}
\multiput(0,80)(0,20){2}{\line(1,0){16}}
\end{picture}%
\begin{picture}(40,100)
\put(0,-5){\framebox(40,90){$F_m$}}
\put(0,100){\line(1,0){40}}
\end{picture}%
\begin{picture}(16,100)
\multiput(0,0)(0,20){2}{\line(1,0){16}}
\multiput(8,44)(0,4){3}{\makebox(0,0){.}}
\multiput(0,80)(0,20){2}{\line(1,0){16}}
\end{picture}%
}
\caption{\label{diederkreis} Quantum circuit for the implementation of
the dihedral POVM.}
\end{figure}

Let $D_{2m}= \langle r,s : r^m=1, s^2=1,srs^{-1}=r^{-1}\rangle$ be the
dihedral group \cite{Jacobson} with $n=2m=2^{k+1}$ elements for a
fixed $m=2^k\geq 4$.  The element $r$ denotes the rotation and $s$ the
reflection of the dihedral group. We consider the irreducible
representation $\varphi:D_{2m}\to {\cal U}(2)$ that is defined
by 
\[
\varphi(r) =\left(\begin{array}{cc} \omega & 0 \\ 0 & \omega^{-1}
\end{array}\right) \quad {\rm and} \quad \varphi(s) = \left(
\begin{array}{cc} 0 & 1 \\ 1 & 0 \end{array} \right).
\]
The element $\omega={\rm exp}(2\pi i /m)$ is an $m$th root of unity.
For $\alpha, \beta \in {\mathbb C}$ with $|\alpha|^2 + |\beta|^2 =
1/m$ we consider the POVM with the corresponding matrix
\[
M = \left(
\begin{array}{cccccc}\alpha &
\ldots & \alpha & \beta & \ldots &\beta \\ \beta & \ldots & \beta
\omega^{m-1} & \alpha & \ldots & \alpha \omega^{m-1} \end{array}
\right) \in {\mathbb C}^{2 \times n}.
\]
The matrix $M \in {\mathbb C}^{2\times n}$ has the symmetry $\varphi M
= M\varphi_{\rm mon}$ where $\varphi_{\rm mon}$ is defined by the
equations $\varphi_{\rm mon}(r) = \onemat_2 \otimes \omega S_m^{-2} $
and $\varphi_{\rm mon}(s) = \sigma_x \otimes F^2_m T_m$. The matrices
$S_m,T_m \in {\mathbb C}^{m \times m}$ are defined by the equations
(indices are taken modulo $m$)
\[
S_m=\sum_{i=0}^{m-1} \ket{i+1}\bra{i}, \quad
T_m=\sum_{i=0}^{m-1} \omega^i \ket{i}\bra{i}
\]
and $F_m$ denotes the discrete Fourier transform defined in the
previous section. In order to decompose $\varphi_{\rm mon}$ into
irreducibles the following permutation $Q_k$ is useful. Denoting by
$\overline{x}$ the binary complement of the binary vector $x$ of
length $k$ we define $Q_k : \ket{x, 0} \mapsto \ket{x, 0}$ and $Q_k:
\ket{x, 1} \mapsto \ket{\overline{x}, 1}$. 
Furthermore, we introduce the representations $\varphi_l$ defined by
\[
\varphi_l(r)=\left( \begin{array}{cc}\omega^l & 0 \\
0&\omega^{-l}\end{array}\right) \quad {\rm and} \quad
\varphi_l(s)=\left( \begin{array}{cc}0 & 1 \\ 1& 0
\end{array}\right).
\] 
With this notation we have $\varphi = \varphi_1$.  The two-dimensional
representations $\varphi_l$ are irreducible and inequivalent
\cite{Jacobson} for different $l \in \{1,\ldots, m/2\}$.
Now, using the base change
$W:= Q_m (\onemat_2 \otimes F_m^\dagger)\in {\mathbb C}^{n \times n}$
we obtain that
\[
W \varphi_{\rm mon} W^\dagger = \psi\oplus \psi \oplus \psi \oplus \psi\,,
\] 
where $\psi$ is a direct sum of all 
representations $\varphi_j$ 
with odd $j$. The first component of $\psi$ is $\varphi_1$, the
other components $\varphi_j$ appear  
in a specific order which is irrelevant in the sequel.
 We choose
the representation 
\[
\varphi^\prime = \psi'\oplus \psi \oplus \psi \oplus \psi\,,
\]
where $\psi'$ is obtained from $\psi$ by dropping $\varphi_1$.
This
leads to $\varphi \oplus \varphi^\prime = W \varphi_{\rm mon}
W^\dagger$. The matrix $C=MW^\dagger = \left(
\sqrt{m}\alpha \, 0 \, \ldots \, 0 | \sqrt{m} \beta \, 0 \, \ldots
\, 0 \right) \otimes \onemat_2 \in {\mathbb C}^{2 \times n}$
defines the first two rows of the intertwining matrix ${\tilde C}$ we
want to construct according to Algorithm \ref{POVMalg} from Section
\ref{POVMorth}.  
A possible extension of the intertwining matrix $C\in {\mathbb C}^{2
\times n}$ to a unitary matrix ${\tilde C} \in {\cal U}(n)$ 
is ${\tilde C}= A \otimes \onemat_{m/2}$ with
the matrix
\[
A = \sqrt{m} \left(
\begin{array}{cc} \alpha & \beta \\ {\overline \beta} & - {\overline
\alpha} \end{array} \right) \in {\cal U}(2).
\]
According to Algorithm \ref{POVMalg} from Section \ref{POVMorth} we
have to define the matrices $U \in {\cal U}(2)$ and $V
\in {\cal U}(n-2)$.  The equations
$\varphi=\varphi_1$ and $W\varphi_{\rm mon}W^\dagger =
(\varphi_1 \oplus \psi') \oplus \psi \oplus \psi \oplus \psi$
show that
$U=\onemat_2$. Furthermore, we choose $V=\onemat_{n-2}$. Then we have
the matrix $U\oplus V= \onemat_n$.  To summarize, we have to implement
the matrix
\[
{\tilde M}^\dagger= W^\dagger {\tilde C}^\dagger=
(\onemat_2 \otimes F_m) Q_k 
(A^\dagger \otimes \onemat_4) \in {\cal U}(n)
\] 
in order to measure the POVM corresponding to the dihedral group
$D_m$. The scheme of the circuit corresponding to ${\tilde M}^\dagger$
is shown in Figure \ref{diederkreis}.

\subsection{Weyl-Heisenberg groups}

In the following we introduce the finite Weyl-Heisenberg groups which are
matrix groups acting on a finite dimensional vector space. For our
purposes we consider vector spaces of dimension $m=2^k$ only, where $k
\geq 2$. Then the Weyl-Heisenberg group $G_m$ is the group generated
by the matrices $S_m=(1, 2, \ldots, m)$ and $T_m={\rm diag} (1,
\omega, \omega^2, \ldots, \omega^{m-1})$ where $\omega ={\rm exp}(2
\pi i/m)\in {\mathbb C}$ is a primitive $m$th root of unity. It is
known that $G_m$ contains $m^3$ elements \cite{Terras:99}. POVMs that
are covariant with respect to the Weyl-Heisenberg groups have a
physical motivation. Since the position and momentum of a particle
cannot be measured simultaneously by any projection-valued measurement
one has to construct POVMs which measure both observables with a
certain inaccuracy.  This idea has already been described in
\cite{Davies2}: starting from a wave packet, i.e., a unit vector 
$|\psi \rangle \in
L^2({\mathbb R})$ we define a set $\{M_{s,t}\}$ of operators by
\[ 
M_{s,t}:= \frac{1}{2\pi}e^{isP+tQ} |\psi \rangle \langle \psi | e^{-isP-tQ}\,.
\]
where $s,t\in \R$ and $P$ and $Q$ are the position and momentum
operators, respectively. Explicitly, they are defined by
$(P\psi)(x):=-i(d/dx)\psi (x)$ and $(Q\psi)(x):=x\psi(x)$.
 We then have that 
\[
\int_{s,t} M_{s,t}\, ds dt =1.
\]
The POVM $\{M_{s,t}\}$ provides an approximative realization of the
classical phase space since the measurement outcome $(s,t)$ can be
interpreted as the point $(s,t)$ in the phase space. In the following
we are interested in finite dimensional approximations of this. Assume
that we want to measure the position and crystal momentum of a
particle on a lattice with $m$ points for $m=2^k$ \cite{Zim}. 
Furthermore, we
assume that it is possible to transfer the state of such a system into
$k$ qubits of a quantum register.  That means that we can implement a
bijection of the basis states with Hamming weight one to the basis
states of the Hilbert space ${\mathbb C}^m$ of the $k$ qubits.  The
canonical basis states $\ket{j}$ of ${\mathbb C}^m$ denote the
position eigenstates.  The states corresponding to the state vectors
$\sum_{j=0}^{m-1}e^{2\pi i l j/m} \ket{j}$
with $l=0,\dots,m-1$ are the eigenstates of the crystal momentum.
Explicitly, the crystal momentum $p$ can be defined by $p:=2\pi l/m
-\pi$. With this definition the values of $p$ are in the interval
$[-\pi,\pi]$ that meets the usual physical intuition of the
one-dimensional Brillouin zone of an infinite one-dimensional
crystal. Here we characterize the position and momentum simply by the
integer values $j, l =0,\dots,m-1$.  The cyclic translation of the
position is given by the action of $S_m$ and a change of crystal
momentum by the action of $T_m$.  Consider a rank-one positive
operator $\ket{\psi}\bra{\psi}$ with the property that neither the
position nor the momentum of the corresponding state is completely
undefined.  Set
\[
M_{j,l}:= \frac{1}{m} S_m^j
T_m^l \ket{\psi}\bra{\psi} T_m^{-l} S_m^{-j}.
\]
Due to irreducible group action 
the equation $\sum_{j,l} M_{j,l} =\onemat_m$ holds and the
operators $M_{j,l}$ define a POVM.  For large $m$ we can find states
with corresponding state vectors $\ket{\psi}$ such that both values
$j$ and $l$ are approximately defined. Here the word ``approximately''
is understood with respect to the cyclic topology, i.\,e., $m-1$ and
$0$ are ``almost'' the same value.  A good choice for the POVM will be
the following.  Set $\ket{\psi}:=\sum_j c_j \ket{j} $ where the
coefficients $c_j$ are chosen such that the function $j\mapsto
|c_j|^2$ has a unique maximum at $j_0$ and the modulus of the values
$c_j$ decrease with increasing distance from $j_0$ in the cyclic
topology.  If all values $c_j$ are real and they decrease not too
quickly the momentum $l$ of the state is around $j_0$, too.  Then the
measurement values $j,l$ can directly be interpreted as a good estimation
for the position
and momentum values.  We will show that an efficient implementation of
the POVM can be found in the case where 
$
\ket{\Psi}=1/\sqrt{\kappa}(1,\alpha, \alpha^2, \ldots, \alpha^{m/2-2},
\alpha^{m/2-1}, \alpha^{m/2-1}, \alpha^{m/2-2}, \ldots,
\alpha^2,\alpha,1)^T \in {\mathbb C}^m
$ 
with $\alpha \in {\mathbb C}$ and an appropriate normalization factor
$1/\sqrt{\kappa}$.

\begin{figure}
  \centerline{\epsffile{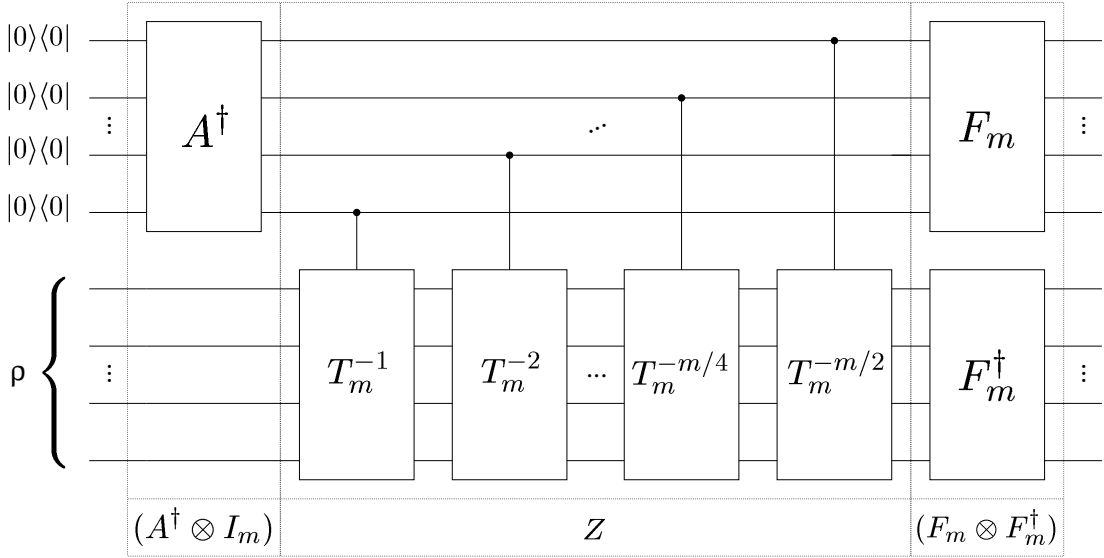}} 
  \caption{Circuit for the implementation of the POVM with respect to 
the Weyl-Heisenberg group and the vector $\ket{\Psi}=(v_1, \ldots, v_m)^T$.
The vector $\ket{\Psi}$ determines the matrix $A^\dagger$.}  
  \label{HWGBild}
\end{figure}

In the following we consider the group-generated POVMs with respect to
$G_m$ and the natural representation $\varphi$ defined by
$\varphi(g)=g$ for all $g\in G_m$. This representation is irreducible.
Therefore, following Algorithm \ref{POVMalg} from Section
\ref{POVMorth} we can set $U=\onemat_m$ since $\onemat_m$ decomposes
$\varphi$ into a direct sum of irreducible representations.  The
vector $\ket{\Psi}=(v_1, \ldots ,v_m)^T \in {\mathbb C}^m$ with the
normalization $|v_1|^2 + \ldots + |v_m|^2 = 1/m$ leads to the POVM
where the defining matrix $M \in {\mathbb C}^{m \times n}$ is
given by 
\[
\left(
\begin{array}{ccccccccccc}
v_1&v_1&\ldots&v_1&v_m&\ldots&v_m&\ldots&v_2&\ldots&v_2 \\ v_2&v_2 \omega&\ldots&v_2
\omega^{m-1}&v_1&\ldots&v_1 \omega^{m-1}
&\ldots&v_3&\ldots&v_3\omega^{m-1}\\
&\vdots&\ddots&&&\ddots&&\ddots&&\ddots\\ v_m&v_m\omega^{m-1}&\ldots&v_m
\omega&v_{m-1}&\ldots&v_{m-1}\omega&\ldots&v_1&\ldots&v_1 \omega
\end{array}\right).
\]
Note that we identify vectors $g\ket{\Psi}$ and $h\ket{\Psi}$ for
different $g, h \in G_m$ that are equal up to a global phase
factor. Consequently, the POVM consists of at most $n=m^2$ different
operators. For example when $m=4$ the vector
$\ket{\Psi}=(v_1,v_2,v_3,v_4)^T \in {\mathbb C}^4$ with
$|v_1|^2+|v_2|^2+|v_3|^2 + |v_4|^2 = 1/4$ leads to the POVM with
$n=16$ operators and the corresponding matrix $M \in {\mathbb C}^{4
\times 16}$ where $M$ is defined by
\[
\left( \begin{array}{rrrrrrrrrrrrr}
v_1&v_1&v_1&v_1&v_4&v_4&v_4&v_4&\ldots&v_2&v_2&v_2&v_2\\
v_2&v_2i&-v_2&-v_2i&v_1&v_1i&-v_1&-v_1i&\ldots&v_3&v_3i&-v_3&-v_3i\\
v_3&-v_3&v_3&-v_3&v_2&-v_2&v_2&-v_2&\ldots&v_4&-v_4&v_4&-v_4\\
v_4&-v_4i&-v_4&v_4i&v_3&-v_3i&-v_3&v_3i&\ldots&v_1&-v_1i&-v_1&v_1i
\end{array}\right).
\]
The symmetry of $M\in {\mathbb C}^{m \times n}$ can be described on
the generators by the equations $T_m M=M(\onemat_m\otimes S_m)$ and
$S_m M=M(S_m\otimes T_m^\dagger)$.  Therefore the representation $\varphi_{\rm
mon}: G_m \to {\cal U}(n)$ is defined by $\varphi_{\rm
mon}(T_m)=\onemat_m \otimes S_m$ and $\varphi_{\rm mon}(S_m) = S_m
\otimes T_m^\dagger$. The symmetry of $M$ can also be written as
\[
T_m M=M (\onemat_m \otimes T_m)^{F_m \otimes F_m^\dagger} \quad {\rm
and} \quad S_mM=M(T_m^\dagger\otimes S_m)^{F_m \otimes F_m^\dagger}
\]
where we use the notation $A^X=XAX^\dagger$ and the Fourier transform
$F_m$ as defined in Section \ref{zyklische Gruppen}. We can write
$(\onemat_m \otimes T_m)$ and $(T_m^\dagger \otimes S_m)$ as direct
sums
\[
(\onemat_m \otimes T_m) = T_m \oplus T_m \oplus \ldots \oplus T_m
\quad {\rm and} \quad (T_m^\dagger \otimes S_m) = S_m \oplus
\omega^{m-1} S_m \oplus \ldots \oplus \omega S_m.
\]
By using the equations $T_m S_m T_m^\dagger = \omega S_m$ and
$(\onemat_m \otimes S_m)^Z=(T_m^\dagger \otimes S_m)$ we can conjugate
these matrices with the diagonal matrix $Z = \onemat_m \oplus
T_m^{m-1} \oplus T_m^{m-2} \oplus \ldots \oplus T_m^2 \oplus T_m$ in
order to obtain the equations
\[
T_m M=M(\onemat_m \otimes T_m)^{(F_m \otimes F_m^\dagger)Z} \quad {\rm
and} \quad S_mM=M(\onemat_m \otimes S_m)^{(F_m \otimes F_m^\dagger)Z}.
\]
These equations show that we have the decomposition $W\varphi_{\rm
mon}W^\dagger =\varphi \oplus \ldots \oplus \varphi$ with the matrix
$W=Z^\dagger (F_m^\dagger \otimes F_m)$. The representation
$W\varphi_{\rm mon}W^\dagger$ contains $m$ components
$\varphi$. Following Algorithm \ref{POVMalg} from Section
\ref{POVMorth} we have to find a representation $\varphi^\prime$ that
leads to the direct sum $\varphi \oplus \varphi^\prime = \varphi
\oplus \ldots \oplus \varphi$ with $m$ components $\varphi$.
Consequently, we choose $\varphi^\prime = \varphi \oplus \ldots \oplus
\varphi$ with $m-1$ components $\varphi$.  We now consider the
extension of the matrix $C=MW^\dagger=M (F_m \otimes F_m^\dagger)Z \in
{\mathbb C}^{m \times n}$ to a unitary matrix ${\tilde C} \in {\cal U}(n)$. 
 The matrix $C$ is an element of the intertwining
space
\[
{\rm Int}( \varphi, \varphi \oplus \ldots \oplus \varphi ) = \left\{
(\alpha_1, \ldots, \alpha_n)\otimes \onemat_m : \alpha_j \in {\mathbb
C} \right\} \subseteq {\mathbb C}^{m \times n}.
\]
More precisely, we have
$C=( ( \sqrt{m}v_1, \ldots , \sqrt{m}v_m )
F_m^\dagger ) \otimes \onemat_m \in {\mathbb C}^{m \times n}$.
For example, with $m=4$ we have the group $G_4=\langle S_4, T_4
\rangle$ with $S_4 = (1,2,3,4)$ and $T_4 = {\rm diag}(1,i,-1,-i)$ that
contains $64$ elements. In this example we have the equation
\[
C= \left(\left( v_1,v_2, v_3,
v_4 \right) \left( \begin{array}{rrrr}1 &1&1&1\\1&-i&-1&i\\1&-1&1&-1\\
1&i&-1&-i\end{array}\right)\right) \otimes \left( \begin{array}{cccc}
1&\cdot&\cdot&\cdot\\\cdot&1&\cdot&\cdot\\\cdot&\cdot&1&\cdot\\\cdot&\cdot&\cdot&1\end{array}\right)
\in {\mathbb C}^{4 \times 16}.
\]
The matrix $C \in {\mathbb C}^{m \times n}$ determines the first $m$
rows of the matrix ${\tilde C}$ we want to construct. The matrix
${\tilde C}$ is a unitary matrix of the intertwining space
\[
{\rm Int}(\varphi\oplus \ldots \oplus \varphi, \varphi \oplus \ldots \oplus
\varphi) =\{A \otimes \onemat_m : A \in {\mathbb C}^{m\times m}\}
\subseteq {\mathbb C}^{n \times n}.
\] 
When we write ${\tilde C}=A\otimes \onemat_m$ then the matrix $C$
determines the first row of $A$. Explicitly, the first row of $A$ is
\begin{equation}\label{vektor}
\left( 
\sqrt{m}v_1, \ldots, \sqrt{m}v_m \right) F_m^\dagger.
\end{equation}
The operation ${\tilde M}^\dagger$ for the implementation of the POVM
is defined by
\[
{\tilde M}^\dagger = W^\dagger {\tilde C}^\dagger (U \oplus V)=
(F_m \otimes F_m^\dagger) Z (A^\dagger \otimes \onemat_m) \in {\cal U}(n).
\] 
In this equation we have $V=\onemat_{n-m}$ leading to $U\oplus
V=\onemat_m \oplus \onemat_{n-m}=\onemat_n$.  The general scheme for
the implementation of the matrix ${\tilde M}^\dagger$ is shown in
Figure \ref{HWGBild}.
For $m=2^k$ the circuit contains the $k$ 
controlled operations 
\[
T_m^{-1}, T_m^{-2}, \ldots, T_m^{-m/4}, T_m^{-m/2}
\] 
for the implementation of the matrix $Z$.  The matrix $T_m={\rm
diag}(1,\omega,\omega^2, \ldots, \omega^{m-1})$ can be written as
Kronecker product
\[
T_m=
\left(\begin{array}{cc}1&0\\0&\omega^{m/2}\end{array}\right) \otimes
\left(\begin{array}{cc}1&0\\0&\omega^{m/4}\end{array}\right)
\otimes \ldots \otimes \left(\begin{array}{cc}1&0\\0&\omega\end{array}\right) \in
{\cal U}(m).
\] 
Therefore, the matrices $T_m^j$ of the circuit in Figure \ref{HWGBild}
can be implemented efficiently on a register of qubits.

\begin{figure}
  \centerline{\epsffile{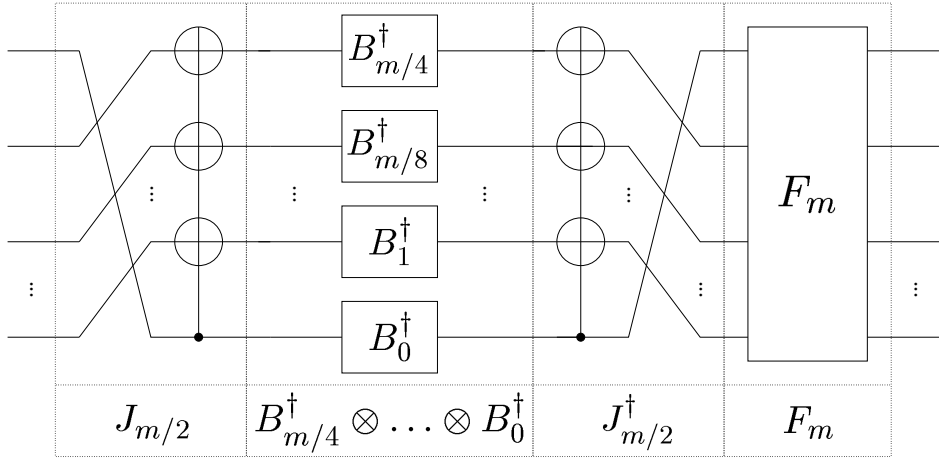}} \caption{Implementation of the
  matrix $A^\dagger$ where $A$ is a matrix that contains the vector
  (\ref{vektor}) as first row. This matrix is part of the circuit in
  Figure \ref{HWGBild} for the vectors (\ref{vektor2}).}
  \label{A-Kreis}
\end{figure}

The circuit in Figure \ref{HWGBild} is efficient if the matrix $A$
that contains the vector (\ref{vektor}) as first row can be
implemented efficiently. We can find such a matrix for the POVM with
the vector
\begin{equation}\label{vektor2}
\ket{\Psi}=\frac{1}{\sqrt{\kappa}}(1,\alpha,\alpha^2, \ldots, \alpha^{m/2-2},
\alpha^{m/2-1}, \alpha^{m/2-1}, \alpha^{m/2-2}, \ldots, \alpha^2,
\alpha,1)^T \in {\mathbb C}^m \end{equation} where we have $\alpha \in
{\mathbb C}$ and the normalization $\kappa=2m(1 + |\alpha|^2 +
|\alpha|^4 + \ldots + |\alpha|^{m-2})$.  A matrix 
$A \in {\cal U}(m)$ that contains the vector (\ref{vektor}) as
first row is given by
\[
A= J_{m/2}^\dagger \left( B_{m/4}\otimes B_{m/8} \otimes \ldots \otimes B_4 
\otimes B_{2} \otimes B_{1} \otimes B_0\right) J_{m/2} F_m^\dagger
\] 
where we use the unitary matrices 
\[
B_{j} = \frac{1}{\sqrt{1+|\alpha|^{2j}}} \left( \begin{array}{cc} 1 &
\alpha^j \\ {\overline \alpha}^j & -1 \end{array} \right) \in {\cal U}(2)\,.
\]
Here $J_k$ is defined to be the permutation matrix which maps $2i
\mapsto i$ and $(2i-1) \mapsto -i$ for $i=0, \ldots, k$. In our
example with $m=4$ we have the matrix
\[
J_{2}^\dagger \left( B_1 \otimes B_0 \right) J_2 =
\frac{1}{\sqrt{2+2|\alpha|^2}}\left(\begin{array}{cccc} 1 & \alpha &
\alpha & 1 \\ {\overline \alpha} & -1 & -1 & {\overline \alpha} \\
{\overline \alpha}&-1&1&-{\overline \alpha} \\ 1 & \alpha & - \alpha &
-1
\end{array} \right).
\]
The circuit scheme for the implementation of the matrix 
\[
A^\dagger = F_m J_{m/2}^\dagger \left( B_{m/4}^\dagger \otimes
B_{m/8}^\dagger \otimes \ldots \otimes B_{4}^\dagger \otimes
B_2^\dagger \otimes B_1^\dagger \otimes B_0^\dagger \right)J_{m/2}
\]
is shown in Figure
\ref{A-Kreis}.

%
%

\section{Conclusions and outlook}

We have shown that a group-covariant POVM can be reduced
to an orthogonal measurements
by a
unitary transform which is symmetric in the sense that it intertwines
two different group representations.
The symmetry of the unitary transform can
be used to derive decompositions which in several cases of interest
(as the Heisenberg-Weyl group)
leads to an efficient quantum circuit for the implementation of the
POVM.

We have argued that POVMs are often necessary in order to understand
why large quantum systems show typically classical
behavior on the phenomenological level.
The POVM with Heisenberg-Weyl symmetry as well
as the example in \cite{SpinAriano} show that the POVMs which
appear in this context are often covariant with respect to some group.

Besides the physical motivation to study implementations of POVMs by
means of orthogonal measurements in terms of quantum circuits there is
also a motivation from computer science. The so-called {\em hidden
subgroup problem} \cite{BH:97} is an attractive generalization of the
quantum algorithms for discrete logarithms and factoring
\cite{Shor:97}. The standard approach for the hidden subgroup problem
consists in a Fourier transform for the respective group followed by a
suitable post-processing on the Fourier coefficients
\cite{HRT:2003}. For abelian groups this post-processing consists
simply in an orthogonal measurement in the computational
basis. However, for non-abelian group measurements which are in fact
POVMs are often more advantageous, see e.~g. \cite{MRRb:2004}. The
POVMs which appear to be useful to solve hidden subgroup problems for
non-abelian groups are naturally group-covariant. The methods
presented in this paper might be useful to find quantum algorithms for
the hidden subgroup problem for new classes of non-abelian groups.

\section*{Acknowledgements}
The authors acknowledge helpful discussions with Markus Grassl. This work was
supported by grants of BMBF project 01/BB01B.  M.~R. has been
supported in part by MITACS and the {\em IQC Quantum Algorithm
Project} funded by NSA, ARDA, and ARO.

%
%


\end{document}